\documentclass[a4paper,fleqn,usenatbib]{mnras}

\usepackage{newtxtext,newtxmath}

\usepackage[T1]{fontenc}
\usepackage{ae,aecompl}
\usepackage[justification=centering]{caption}

\usepackage{graphicx}	
\usepackage{amsmath}	

\newcommand{\snhcc}{SN\,2017hcc}
\newcommand{\flux}{erg\,cm$^{-2}$\,s$^{-1}$} 
\newcommand{\countrate}{counts\,s$^{-1}$}
\newcommand{\columndensity}{cm$^{-2}$ }
\newcommand{\kms}{km s$^{-1}$ }

\newcommand\luminosity{erg\,s$^{-1}$}
\newcommand\spectralluminosity{erg\,s$^{-1}$\,Hz$^{-1}$}
\newcommand\massloss{$M_\odot$\,yr$^{-1}$}

\newcommand\chandra{{\em Chandra}}

\newcommand\spitzer{{\em Spitzer}}
\newcommand\swift{{\em Swift}-XRT}
\newcommand{\comment}[1]{}

\title[Multiwavelength  study of  \snhcc]{THE LUMINOUS TYPE IIN SUPERNOVA SN 2017hcc: INFRARED BRIGHT, X-RAY AND RADIO FAINT}

\author[P. Chandra et al.]{
Poonam Chandra,$^{1, 2}$\thanks{E-mail: pchandra@nrao.edu}
Roger A. Chevalier,$^{3}$
 Nicholas J. H. James$^{3}$
and Ori D. Fox$^{4}$
\\
$^{1}$National Radio Astronomy Observatory, 520 Edgemont Road, Charlottesville, VA 22903, USA\\
$^{2}$National Centre for Radio Astrophysics, Tata Institute of Fundamental Research, Ganeshkhind , Pune 411007, India\\
$^{3}$Department of Astronomy, University of Virginia, P.O. Box 400325, Charlottesville, VA 22904-4325, USA\\
$^{4}$Space Telescope Science Institute,  3700 San Martin Drive,   Baltimore, MD 21218, USA
}
\date{Accepted XXX. Received YYY; in original form ZZZ}

\pubyear{2022}

\begin{document}
\label{firstpage}
\pagerange{\pageref{firstpage}--\pageref{lastpage}}
\maketitle

\begin{abstract}
We present multiwavelength observations of supernova (SN) 2017hcc with the {\em Chandra} X-ray telescope and the X-ray telescope onboard {\em Swift}  ({\em Swift-XRT}) in X-ray bands, with the Spitzer and the TripleSpec spectrometer in near-infrared (IR) and mid-IR bands and with the Karl G. Jansky Very Large Array (VLA) for radio bands. The X-ray observations cover a period of 29 to 1310 days, with the first X-ray detection on day 727 with the {\em Chandra}. The SN was subsequently detected in the VLA radio bands from day 1000 onwards. While the radio data are sparse, synchrotron-self absorption is clearly ruled out as the radio absorption mechanism. The near- and the mid-IR observations showed that late time IR emission dominates the spectral energy distribution. 
The early properties of \snhcc\ are consistent with shock breakout into a dense mass-loss region,  with $\dot M \sim 0.1$ \massloss\ for a decade.
At few 100 days,  the mass-loss rate declined to  $\sim 0.02$ \massloss, as determined from the dominant IR luminosity.
In addition, radio data also allowed us to calculate a mass-loss rate at around day 1000,
 which is two orders of magnitude smaller than the mass-loss rate estimates around the 
bolometric peak. These values indicate that  the SN progenitor  underwent an enhanced mass-loss event a decade before the explosion.
The high ratio of IR to X-ray luminosity is not expected in simple models and is possible evidence for an asymmetric circumstellar region.

\end{abstract}




\section{Introduction}  \label{sec:intro}

The IIn class of supernovae (SNe IIn) was identified by \citet{schlegel90} based on their narrow optical emission lines and hot continua.
The narrow lines can be attributed to slow-moving circumstellar  (CS) matter around the supernova (SN) and the
luminosity can be identified with power provided by  shock waves.
If the cooling time for the gas is fast compared to the age, the SN power is efficiently converted to radiation in this class of
supernovae (SNe).
The required density of the circumstellar medium (CSM) is  high in several SNe IIn \citep[e.g.][]{fransson+14} and are generally
 not seen in Galactic sources  outside of Luminous Blue Variables (LBVs) during  outburst \citep{so06,vm08}.  Radioactivity is not a suitable power source for these SNe because it predicts a faster decline than observed.

In general, the first electromagnetic signal in SNe occurs when the shock reaches the SN photosphere and breaks out on time scales of minutes to hours. 
However, in  dense winds shock breakout can happen in the CSM and last for an extended duration.  There is increasing evidence that in some SNe IIn, the early light curves can be powered by shock  breakouts \citep{ci11, ofek+14, wk17}.  

The early optical spectra of  SNe IIn
 often show emission lines from a wind  that are broadened by electron scattering \citep{chugai01}.
Photons from the inner part of the mass-loss region scatter as they leave the mass-loss region.
An electron column optical depth of about a few, or $10^{24}$ electrons cm$^{-2}$ column density, is needed to be consistent with the observations.
At later times ($> 100$ days), the line profiles are no longer consistent with electron scattering.  Profiles are often  fit
by multiple Gaussians  \citep[e.g.,][]{kiewe+12, szalai+21}. 
Although reasonable fits can be obtained, there is not a clear theory that points to Gaussian line shapes. 
Velocities corresponding to the line widths are larger than the thermal velocities in the optically emitting gas.

Considering the fast shock waves moving into a dense CSM, strong X-ray emission can be expected.
Indeed, SNe IIn are  well represented among the most X-ray luminous SNe \citep{dwarkadas+12,chandra18}.
However, there are optically  luminous SNe IIn that are weak X-ray emitters; a prime example 
is SN 2006gy \citep{ofek+07, smith+07}.
In some cases, the paucity of X-rays can be explained by the effects of high X-ray absorption due to the large column
density through the CS matter \citep{ci12,svirski+12}.
Low X-ray luminosity can also be attributed to a relatively  low density CSM, as in the case of SN 1998S \citep{pooley+02}.
\citet{ci11} have shown that in some  SNe high optical luminosity and low X-ray luminosity can be explained by shock breakout in dense surroundings.

SNe IIn can also be radio emitters from synchrotron radiation by shock accelerated electrons.
Many high luminosity  SNe IIn are also found to be weak emitters in radio bands  \citep{chandra+15}.
 While one expects a high intrinsic radio luminosity owing to a dense CSM, the same high 
density CSM could lead to  efficient absorption of radio emission, e.g.,
the bright Type IIn SN 2010jl became radio bright only after 500 days \citep{chandra+15}.
Thus  radio emission in SNe IIn is likely to be an interplay between the two.   
Internal free-free absorption can also play an important role in these SNe due to efficient mixing of cool gas between forward  and reverse shock   
into the synchrotron emitting region
 \citep{weiler+90, chandra+12b,chandra+20}.
 
   \cite{gerardy+02} first described the late bright infrared (IR) emission from SNe IIn in which the IR light dominates the spectral energy distribution (SED).
 Such emission has been observed in many SNe IIn, 
 including SN 1995N \citep{vandyk13}, SN 1998S \citep{gerardy+00, fassia+00,ms12}, SN 2006jd \citep{fox+11,stritzinger+12},  SN 2007rt \citep{trundle+09,szalai+21},  SN 2010jl \citep{fransson+14, bevan+20},  SN 2005ip \citep{fox+09,fox+10,stritzinger+12, bak+18}, SN 2013L \citep{andrews+17} etc., 
 beginning at an age of $\sim 1$ yr.
The IR emission in these SNe has an approximately blackbody distribution, with a temperature $T$ in the range $\sim 700 - 1800$ K, which suggests that  the emission is from dust.

 There are two uncertainties about the dust emission:  where is it located and how is it powered.
 \cite{gerardy+02} and \citet{fox+11} argue that the IR emission is from pre-existing dust that is beyond the evaporation radius and  the forward shock front.
 In this case,
a decline of the IR emission is expected once the forward shock has overrun the dust shell, around $3-4$ years after the explosion.
Such a trend has been seen in some SNe IIn \citep{fox+11}.      
 On the other hand, efficient dust formation has been considered to be a viable scenario in
 interacting SNe,   as the dust can form  rapidly in the extremely
dense, post-shock cooling layers \citep{smith14}. 
\citet{gall+14} and \citet{sarangi+18}  propose that the dust in SN 2010jl formed in the cold dense shell (CDS)
 downstream from the radiative shock wave.
 They find that the dust can form at an age of about 1 year, when there is an increase in the IR emission from the SN.
 At earlier times, the radiation field is too strong to allow dust formation.
 In addition, dust may also form in the inner SN ejecta. Signatures of the dust formation were reported in SN 2006jc
\citep{smith+08b}, SN 2005ip \citep{smith+09} and SN 2010bt \citep{rosa+18}. The main signatures of dust formation   are an IR
excess, a drop in optical brightness via 
dust extinction, and 
asymmetry of the emission-line profiles due to dust  attenuation \citep{smith14}.

While the most straightforward source for powering the late IR emission in SNe IIn would be continuing CS
 interaction,  for SN 2010jl the evolution of the X-ray absorption column implies a drop in the CSM density
 at an age of about 1 year \citep{chandra+15, sarangi+18, dwek+21}.
 The density drop would result in a drop in the power produced at the forward shock.
To alleviate this \citet{sarangi+18} suggested that the observed power is produced at the reverse shock  wave,  but the issue is unsettled.

\snhcc\ (a.k.a ATLAS17lsn) was discovered on  2017 Oct. 2.38 UT (MJD 58028.38) by the Asteroid Terrestrial--impact Last Alert System  \citep[ATLAS, ][]{tonry11}
and classified as Type IIn SN \citep{prieto+17}.
 The SN  reached a peak at  13.7 mag in around $40-45$ days, indicating an absolute peak magnitude of $-20.7$ mag at a distance $\sim 75$ Mpc  \citep{prieto+17}.  The supernova was not detected 
on 2017 September 30.4  at a limiting magnitude of 19.04\footnote{\url{https://wis-tns.weizmann.ac.il/object/2017hcc/discovery-cert}} mag, suggesting it was discovered soon after the explosion. We assume 2017 Oct 1 UT (MJD 58027) as
the date of explosion throughout this paper.

\citet{prieto+17} 
obtained a bolometric light curve  by using a blackbody function to fit the SED  using the data from  ASAS-SN and the ultraviolet telescope (UVOT) onboard {\it Swift}. They obtained a   peak  luminosity of $L_{\rm bol,peak} = (1.34 \pm 0.14) \times 10^{44}$ \luminosity,  making  \snhcc\ one of the most luminous  SNe IIn ever \citep[e.g.,][]{smith+07, ofek+14}.
They constrain the peak risetime to be  $\sim 27$ days and the total radiated energy $\sim 4\times {10}^{50}$ erg.
Early radio observations at 1.4 GHz with the  Giant Metrewave Radio Telescope resulted in a non-detection 
\citep{nc17}.

A very high degree of intrinsic polarization at optical wavelengths was detected from  \snhcc\  ($>4.8$\%)  \citep{mauerhan+17, kumar+19}. This is  the strongest continuum polarization ever reported for a SN.      \citet{kumar+19} noted a 3.5\% change in the Stokes V polarization in 2 months suggesting  a substantial variation in the degree of asymmetry in either the ejecta and/or the surrounding medium of \snhcc. They  also estimated a mass-loss rate of $\dot M=0.12$\,\massloss\  (for a wind speed of 20 km\, s$^{-1}$), which is comparable to the mass-loss rate for an LBV in eruption.
However, based on echelle spectra,  \citet{sa20} 
 suggested the CSM wind to be flowing axi-symmetrically with wind speeds of 
$v_w =  40-50$ \kms  indicating bipolar geometry of the CSM created by losing 10 $M_\odot$ of stellar mass to eruptive  mass ejections.  
\citet{sa20}   modeled the optical and near IR emission of \snhcc\ and found indications  that the SN ejecta were hidden behind the 
 photosphere until day 75. 
They found that the early time symmetric   
profiles changed to asymmetric blueshifted profiles at late times.
They  suggested 
 the asymmetry to be time and wavelength dependent, causing a systematic blueshift in the line profiles.  This was interpreted 
as a signature of the formation of new dust in the SN ejecta as well as in the post-shock gas within the CDS formed between the forward and the 
reverse shocks.
    
 We carried out early to late time observations of \snhcc\ with  the \chandra\ and the  X-ray telescope (XRT) onboard 
the Neil Gehrels Swift Observatory  (\swift) telescopes in X-ray bands,  
with the Karl G. Jansky Very Large Array (VLA)  in
radio bands, and  with    \spitzer\ and the  TripleSpec in IR bands.  In this paper we report multiwaveband  observations  
of \snhcc\ and their interpretation.  The observations are described in \S \ref{sec:obs}. 
 Our main results are summarised in  \S \ref{sec:results}, and the discussion and comparison with published data are in \S \ref{sec:discussion}. 
Finally, our main conclusions are summarised in \S \ref{sec:conclusions}.

\begin{figure*}
\begin{center}
\includegraphics[width=0.43\textwidth,angle=0]{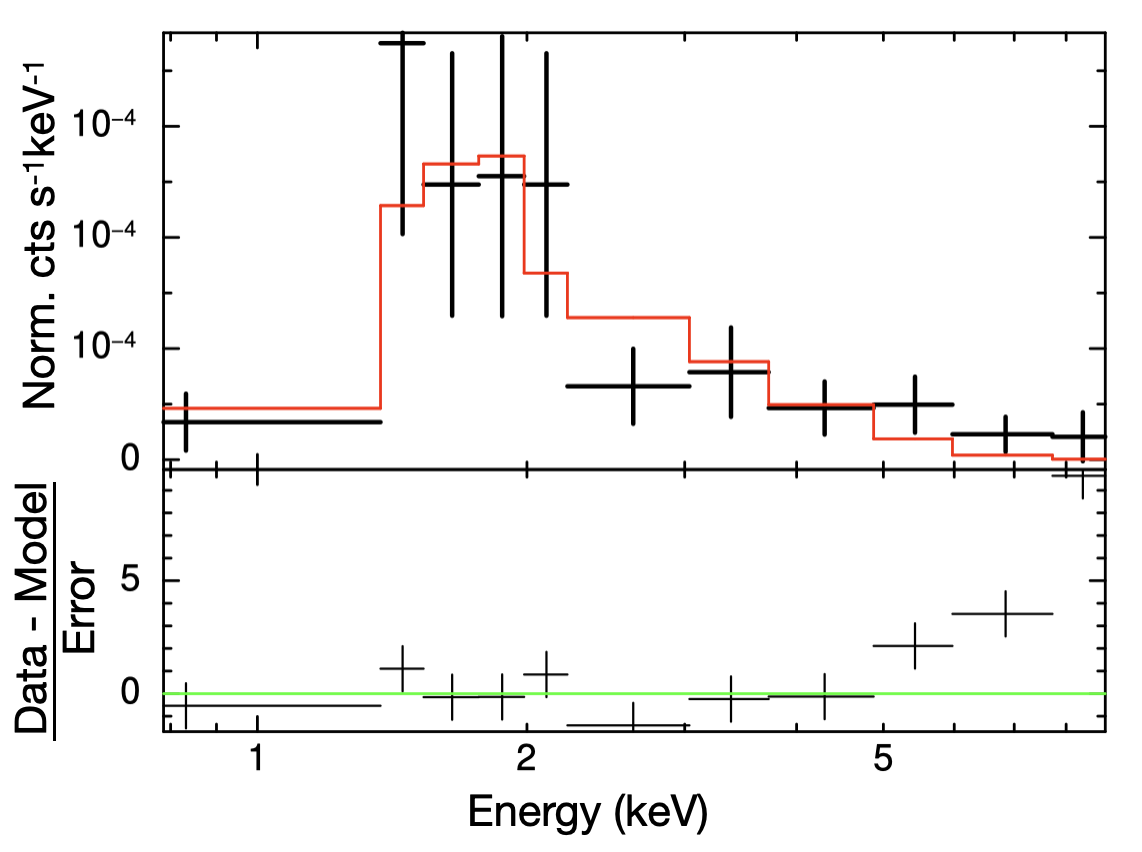}
\includegraphics[width=0.43\textwidth,angle=0]{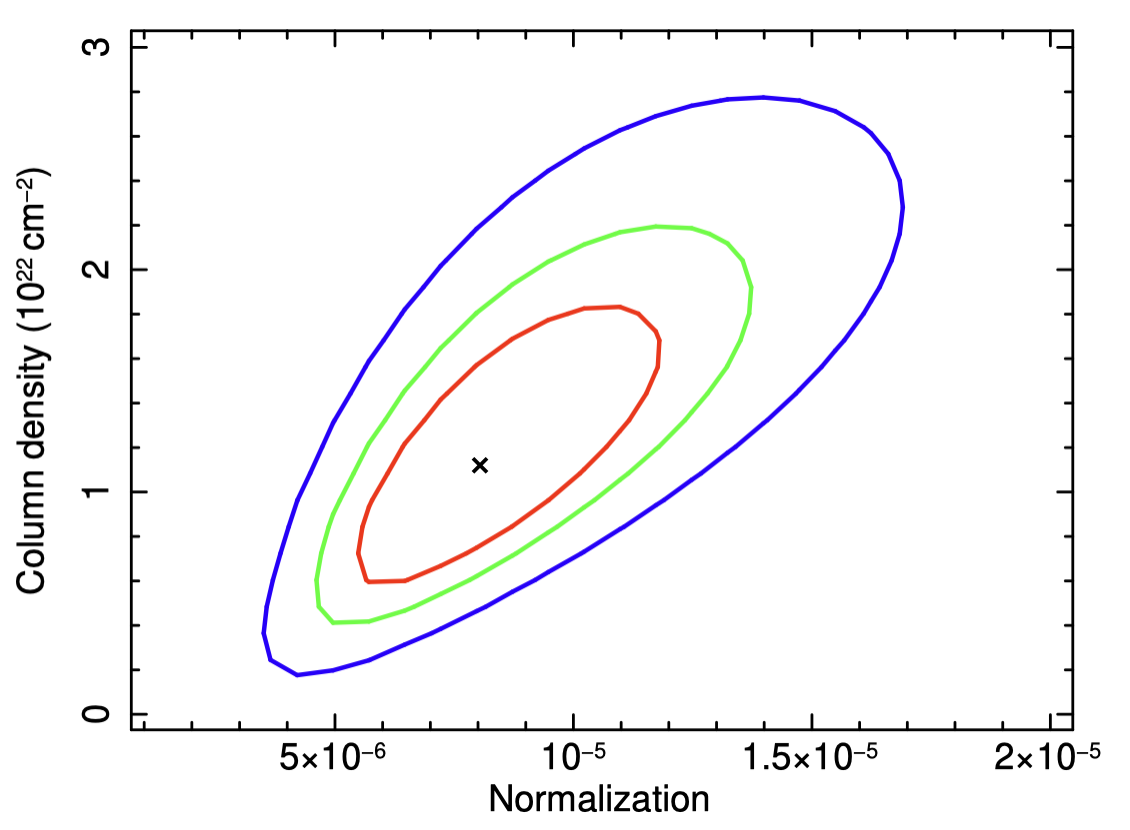}
\caption{\small {\it Left panel:} {\it Chandra} ACIS-S spectrum of \snhcc.  The spectrum is fit with a thermal bremsstrahlung spectrum assuming a temperature of 3 keV. {\it Right panel:}
Contours of best fit column density with the best fit normalization. The red, green and blue lines represent 1-$\sigma$, 
2-$\sigma$ and 3-$\sigma$ contours.
 \label{fig:xray}}
\end{center}
\end{figure*}

 \section{Observations}
 \label{sec:obs}
 
 \subsection{Chandra Observations}

We observed \snhcc\, with the  \chandra\ ACIS-S on 2019 Sep 27  (MJD  58753) for an exposure of 40 ks.
We extracted  the spectrum, response and ancillary matrices using 
Chandra Interactive Analysis of Observations software \citep[CIAO;][]{fruscione+06}, using task
\texttt{specextractor}. The 
CIAO version 4.6 along with CALDB version 4.5.9 was used for this
purpose.  The  HEAsoft\footnote{\url{http://heasarc.gsfc.nasa.gov/docs/software/lheasoft/}}
package Xspec version 12.1 \citep{arnaud+96} was used to carry out the spectral analysis. 
 Due to low counts, only 5 channels were averaged and 
we used  maximum likelihood statistics for a Poisson distribution, i.e.  
the $c$-statistics   \citep{cash79}.  

The spectrum was fit with an  absorbed thermal bremsstrahlung model. 
In the view of \citet{sa20}, the optical measurements implied a  SN ejecta velocity of 
4000 \kms\ around the \chandra\ epoch and a  CDS velocity of 1600 \kms. Assuming the CDS velocity to be that   of the  forward shock, 
the  reverse shock velocity is 2400  \kms\   \citep{sa20}. 
However, in SNe IIn  the CDS generally 
absorbs any X-ray emission coming from the reverse shock due to the high column density of cool gas, and the dominant emission is likely to come from the forward shock \citep{margutti+14, ofek+14, chandra+15}. 
The forward shock velocity of 1600 \kms\ estimated by \citet{sa20} corresponds to a temperature of 
 $\sim 3$ keV, which is what we initially assume to
fit the \chandra\ spectrum.
Our observation resulted in a detection with an unabsorbed $0.3 - 10$\,keV luminosity   of $1.94\times 10^{40}$ \luminosity\ .
The best fit column density is $ 1.13^{+0.78}_{-0.58} \times 10^{22}$ \columndensity\ (Fig. \ref{fig:xray}).

We, however, note that the forward shock velocity could be larger than the above value,  as discussed in \S \ref{sec:discussion1}. 
To reflect this possibility, we also ran the fits 
by fixing the  X-ray emitting shocked plasma temperature to be 20 keV (corresponding to  a  forward shock velocity of 4000\,\kms,  an estimate based on observations of SN\,2010jl at an age of  737 days  \citep{ofek+14, chandra+15}). In this case, our observation resulted in  an unabsorbed $0.3-10$\,keV luminosity   of $1.74\times 10^{40}$ \luminosity\ .
The best fit column density is $ (5.7^{+6.2}_{-4.5}) \times 10^{21}$ \columndensity. 
 While the X-ray luminosity is lower by only 10\% for 20\,keV plasma as compared to that with 3\,keV plasma, the best fit column density is lower by a factor of 2, though we note that the  uncertainty  in column density is
large with a 3-$\sigma$ upper limit
of $\sim 2 \times 10^{22}$ \columndensity. 
Hence this is  roughly consistent with the previous value.

\subsection{Swift-XRT Observations}

 We observed \snhcc\ with the \swift\ starting 2017 Oct 28 (MJD 58054).  The observations continued until 2021 May 03 (MJD 59337) at various
epochs.   The measurements were taken in the photon counting mode.
 The \texttt{xselect} program of the
HEAsoft 
package  (version 6.28, CALDB version (XRT(20210915))) was used to create the spectra and images. Observations closely spaced in time were combined. 
None of the {\it Swift}-XRT observations resulted in detection with 3-$\sigma$ upper limits ranging around  $\sim (1-4)\times 10^{-3}$ \countrate. The $0.3 -10.0$\,keV flux  was obtained from  the $3\sigma$ upper limits on the count rates by assuming a thermal plasma of 3 keV 
and a column density of $1.1\times10^{22}$\,\columndensity. 
The details are given in Table \ref{tab:xray}.

\begin{table*}
\centering
\caption{X-ray observations of SN 2017hcc}
  \label{tab:xray}
\begin{tabular}{llcccrcr}
\hline
Date of & MJD & Telescope &    Exposure &   Age & Count rate &  Unabs. Flux & 
Unabs. Luminosity  \\
Observation (UT)  & &  & 		(ks) &		(day) &  (cts s$^{-1}$) & (erg s$^{-1}$ cm$^{-2}$) &(erg s$^{-1}$)\\
\hline	    
$\sim$ 2017 Oct 30 & 58056 & {\it Swift}-XRT  &  13.47 & 29 & $<1.02 \times10^{-3}$  & $<0.54\times10^{-13}$ & 
$<3.64\times10^{40}$ \\
$\sim$ 2017 Nov 17 &58074 & {\it Swift}-XRT &   5.20 & 47 & $<2.56\times10^{-3}$  & $<1.35\times10^{-13}$ & 
$<8.97\times10^{40}$\\
 $\sim$ 2017 Dec  10 & 58097  & {\it Swift}-XRT & 9.58& 70 & $<1.03\times10^{-3}$  & $<0.55\times10^{-13}$ & 
$<3.65\times10^{40}$ \\
2018 May 23 & 58261  &  {\it Swift}-XRT & 4.94 & 235 & $<3.85\times10^{-3}$  & $<2.05\times10^{-13}$  & 
$<13.7\times10^{40}$\\
2019 Sep 27 & 58753 & {\it Chandra} ACIS-S & 40.53 & 727 & $(1.28\pm0.36)\times10^{-3}$  & $(2.87\pm0.53)\times10^{-14}$  & 
$(1.94\pm0.43)\times10^{40}$\\
2021 May 03 & 59337 &  {\it Swift}-XRT & 4.71 & 1310 & $<2.81\times10^{-3}$  & $<1.48\times10^{-13}$  & 
$<10.0\times10^{40}$\\
\hline
\end{tabular}
 \vspace{1ex}

  {\raggedright  For \chandra\ observations, the temperature was fixed to 3 keV (see \S \ref{sec:discussion}).  The best fit column density is
$N_H=(1.13^{+0.78}_{-0.58}) \times10^{22}$ cm$^{-1}$.
The count rates have been converted to fluxes assuming a thermal plasma of 3 keV and column density of
$1.1\times10^{22}$ cm$^{-1}$ for the \swift\ observations. \par}
\end{table*}

\subsection{Spitzer observations}
\label{sec:spitzer}

\begin{figure}
\begin{center}
\includegraphics[width=0.49\textwidth]{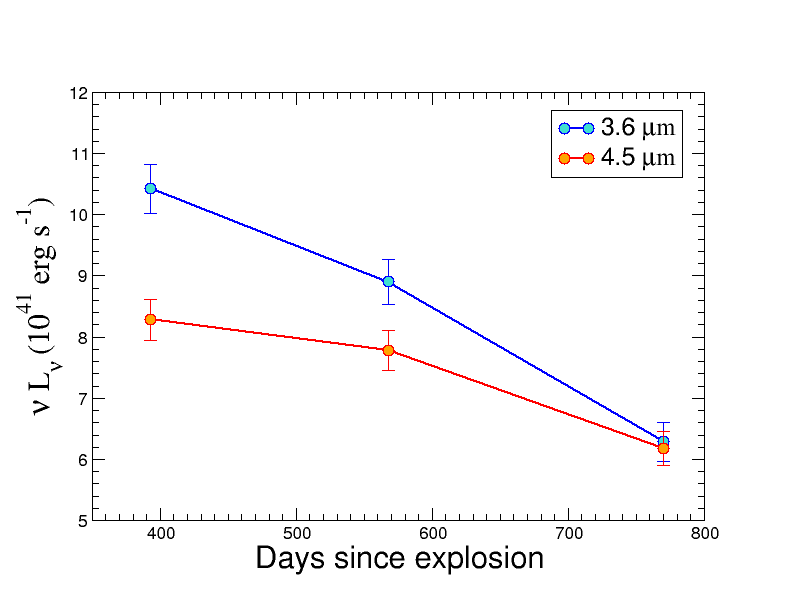}
\caption{\small {\it Spitzer} photometry of SN 2017hcc. The data are taken at 3 epochs (2018 Oct 29 (day 393), 2019 Apr 21 (day 568) and 2019 Nov 9 (day 770))  at 3.6 and 4.5 $\mu$m wavelengths (\S \ref{sec:spitzer}). The temporal evolution indices
at 3.6$ \mu$m between days 393 and 568, and days 568 and 770 are 
$-0.43\pm0.15$, and $-1.14\pm0.22$, respectively.
The temporal evolution indices
at 4.5$ \mu$m between days 393 and 568, and days 568 and 770 are 
$-0.17\pm0.16$, and $-0.76\pm0.20$, respectively. 
\label{fig:spitzer}}
\end{center}
\end{figure}

\snhcc\, was observed with the Infrared Array Camera (IRAC) \citep{fazio+04} onboard \spitzer\ \citep{werner+04}
at three epochs (PI: Fox), starting 2018 Oct 29 (MJD 58420; Table \ref{tab:ir}).  We used  the Spitzer Heritage Archive (SHA)\footnote{https://sha.ipac.caltech.edu} 
to download Post Basic Calibrated Data (pbcd), which are already fully coadded and calibrated.  Standard aperture photometry was performed, although separate apertures were strategically placed by eye to best estimate the local background.  Figure \ref{fig:spitzer} plots and Table \ref{tab:ir} lists the mid-IR photometric data, which are also included in a comprehensive {\it Spitzer} paper by \citet{szalai+21}.              

As in the analysis laid out by \citet{fox+11}, the mid-IR photometry can be fit as a function of the dust temperature, $T_d$, and mass, $M_d$.  Given only two photometry points per epoch, we assume a  simple blackbody model, consisting of a single graphite grain of size  1 $\mu$m and a single temperature.  We calculate opacities using  
the optical constants from \citet{dl84}.
Table \ref{tab:ir} lists the results and Figure \ref{fig:spitzer} shows the \spitzer\ data.

\begin{table*}
\centering
\captionsetup{justification=centering}
\caption{Spitzer IRAC observations of \snhcc}
 \label{tab:ir}
\begin{tabular}{lcrrr}
\hline
Quantity &Wavelength &  \multicolumn{3}{c}{Date of observations}\\
&  & 2018 Oct 29.35 & 2019  Apr 20.58 & 2019 Nov 08.57\\
& & (MJD 58420.35) & (MJD 58593.58) & (MJD 58795.57)\\
\hline
AOR id & & 66120960   & 68799232  & 68799488\\
\hline
Epoch (d) & & 393& 568 & 770\\
\hline
Apparent magnitude & 3.6$\mu$m & 12.97$\pm$0.04 & 13.16$\pm$0.05 &
 13.52$\pm$0.05 \\
  & 4.5$\mu$m & 12.49$\pm$0.04 & 12.56$\pm$0.04 & 12.81$\pm$0.04 \\
 \hline
 {Absolute magnitude} & 3.6$\mu$m & -21.32$\pm$0.19 & -21.14$\pm$0.19 & -20.77$\pm$0.19\\
& 4.5$\mu$m & -21.80$\pm$0.19 & -21.73$\pm$0.19 & -21.48$\pm$0.19\\
 \hline
 {Flux density ($\mu$Jy)} & 3.6$\mu$m & $1818.72\pm70.35$ & $1533.79\pm64.98$ & $1097.13\pm55.58$ \\
& 4.5$\mu$m & $1813.78\pm71.83$ &  $1703.51\pm69.03$ & $1354.05\pm62.36$\\
 \hline 
 {$\nu L_\nu$ (erg s$^{-1}$)} & 3.6$\mu$m & $(10.42\pm0.40)\times10^{41}$  &
 $(8.90\pm0.37)\times10^{41}$  & $(6.29\pm0.32)\times10^{41}$ \\
 {  } & 4.5$\mu$m & $(8.28\pm0.33)\times10^{41}$ & $(7.78\pm0.32)\times10^{41}$
 & $(6.18\pm0.28)\times10^{41}$\\
 \hline
 Dust mass$^a$ ($M_\odot$) & & $2.26\times 10^{-3}$  & $3.60\times 10^{-3}$  & $ 4.57\times 10^{-3}$\\
 \hline
 Dust Temp (K) &  &  $1.28\times 10^{3}$  & $1..07\times 10^{3}$  & $0.93\times 10^{3}$\\
  \hline
 $L_{\rm bol}$$^b$ (erg s$^{-1}$)   &  & $4.80\times 10^{41}$  & $3.73\times 10^{41}$  & $2.70\times 10^{41}$\\
  \hline
 Blackbody radius (cm) &  &  $1.57\times 10^{16}$  & $1.97\times 10^{16}$  & $2.23\times 10^{16}$\\
 \hline
 \end{tabular}
 \vspace{1ex}

      {\raggedright $^a$ The dust mass estimates are sensitive to the chosen grain parameters. The listed values are estimated assuming  1 $\mu$m grain size of graphite composition.   \par}
      {\raggedright $^b$ Bolometric luminosity assuming a single temperature dust blackbody radiation. \par}
\end{table*}

\begin{figure*}
\begin{center}
\includegraphics[width=0.66\textwidth,angle=0]{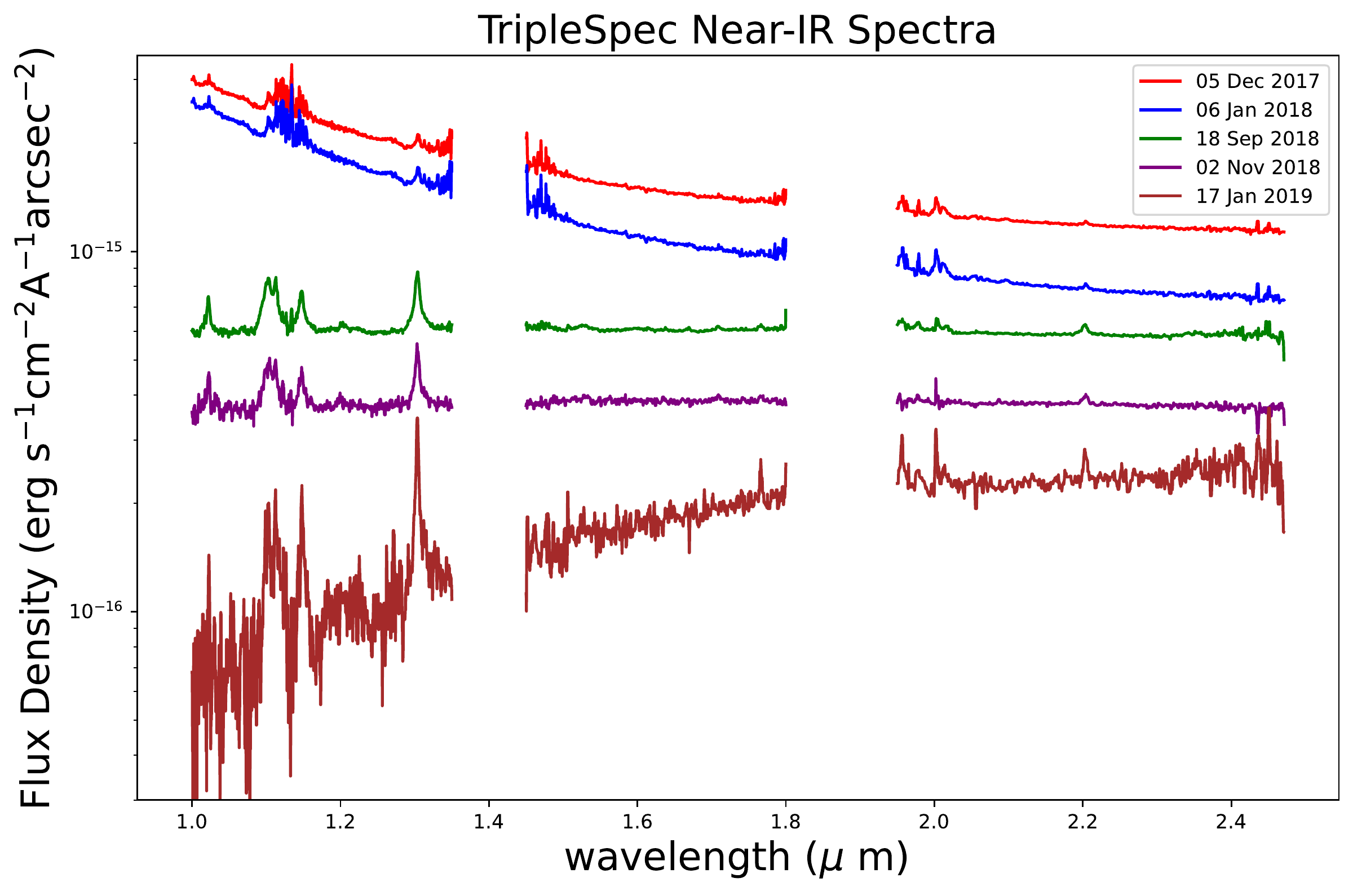}
\caption{ Triplespec near-IR spectra of \snhcc\ on 2017 Dec 05 (day 66), 2018 Jan 06 (day 98), 2018 Sep 18 (day 353), 2018 Nov 02 (day 398) and 2019 Jan 17 (day 474). Note that the early spectra are
brighter at lower wavelengths, whereas the late time spectra are brighter at longer wavelengths. 
To show this trend better, the top four spectra are offset by $10^{-15}$, 
$6\times 10^{-16}$,  $5 \times 10^{-16}$, and $3\times 10^{-16}$ erg\,s$^{-1}$\,cm$^{-2}$\,A$^{-1}$\,arcsec$^{-2}$, respectively.  \label{fig:triplespec}}
\end{center}
\end{figure*}

\subsection{TripleSpec Observations}

 We obtained five epochs,  2017 Dec 05 (MJD 58092), 2018 Jan 6 (MJD 58124), Sep 18 (MJD 58379), Nov 02 (MJD 58424) and 2019 Jan 17 (MJD 58500), of near-IR spectroscopy, with simultaneous continuous wavelength coverage from 0.95-2.46 $\mu$m in five spectral orders, with the TripleSpec spectrograph at the Apache Point Observatory 3.5-m \citep{wilson+04}.  Our observation sequence consists of 300-s dithered exposures that could be pair subtracted to allow for correction of thermal background and night-sky emission lines.  We used a modified version of SpexTool for the spectral extraction \citep{cushing+04}.  While the early data are dominated by shorter wavelength emission, the later data reveal higher flux densities at longer wavelengths (Fig. \ref{fig:triplespec}). However, we note a caveat here that  due to calibration uncertainty  the TripleSpec flux should not be considered absolute and ideally should be scaled to near-IR photometry.

\subsection{VLA observations}

The VLA  observed \snhcc\ between 2020 June 29--30 (MJD 59029--59030) in bands X (8--12\,GHz) and K (18--26\,GHz), and later in June 2021 in Ka (26--40\,GHz), Ku (12--18\, GHz), X, C (4--8\,GHz) and S (2--4\,GHz) bands. The data were analysed 
using the standard packages within the Common Astronomy Software Applications package \citep[CASA, ][]{casa}.
The details of the observations and the flux density values can be found in Table \ref{tab:radio}.
We add 10\% error in the quadrature to the flux density values for analysis purposes, a typical uncertainty in the flux density calibration scale at the observed frequencies
\footnote{https://science.nrao.edu/facilities/vla/docs/manuals/oss/performance/fdscale}.

\begin{table*}
\centering
\caption{Radio observations of \snhcc}
 \label{tab:radio}
\begin{tabular}{llccclccc}
\hline
 Date of   & MJD &  Telescope &     Age & Central & Flux density &   rms & Luminosity &
 $\nu L_\nu$  \\
Obs. (UT) & &		&  (days)  & 		Freq (GHz)	 & $\mu$Jy	& 		$\mu$Jy	 & 
 (erg s$^{-1}$ Hz$^{-1}$) &(erg s$^{-1}$)\\
\hline
2020 Jun 29 & 59029 &VLA &   1002 & 10.0 & 32.5$\pm$7.6 &6.7 & $(2.19\pm0.51) \times10^{26}$  & $(2.19\pm0.51) \times10^{36}$ \\
2020 Jun 30 &  59030 & 	VLA &  1003 & 22.0 & 77.1$\pm$16.2 & 15.5 & $(5.19\pm1.09) \times10^{26}$ & $(11.42\pm2.40) \times10^{36}$ \\
2021 Jun 04 & 59369  & VLA &  1342 & 10.0 & $<45.6$ & 15.26 &  $<3.07 \times10^{26}$ & $<3.07\times10^{36}$\\
2021 Jun 09 & 59374&  VLA &  1347 & 15.4 & 31.0$\pm$11.3 & 10.5 & $(2.09\pm0.76) \times10^{26}$ & $(3.21\pm1.17) \times10^{36}$\\
2021 Jun 09 &  59374 & VLA &  1347 & 6.1 & $<26.7$ & 8.9 & $<1.80 \times10^{26}$ & $<1.10\times10^{36}$\\
2021 Jun 11 & 59376 & VLA &  1349& 3.0 & $<182.4$ & 60.7&$<12.28 \times10^{26}$ & $<3.68 \times10^{36}$ \\
2021 Jun 14 & 59379 & VLA & 1352  & 33.1 & 33.5$\pm$11.8 & 11.6 & $(2.26 \pm0.79) \times10^{26}$ & $(7.46\pm2.63) \times10^{36}$\\
2021 Jun 16 & 59381 & VLA & 1354  & 1.5 & $< 228$ & 76.1 & $< 15.34 \times10^{26}$ & $<2.30 \times10^{36}$\\
\hline
\end{tabular}
\end{table*}

\section{Analysis and Results}
\label{sec:results}

\subsection{X-ray analysis and the column density}

\swift\ observations of \snhcc\ started when the SN was around a month old;  however, 
the first X-ray detection of \snhcc\ was  at an age of $\sim 2$  years with  the \chandra\ telescope.
Our  \chandra\ observations did not have enough counts to determine the shock temperature and the column density separately. We assumed  a temperature of $\sim 3$ keV, corresponding to a CDS velocity of 1600 \kms, as advocated by   \citet{sa20}.
The best fit column density is  $ 1.13^{+0.78}_{-0.58} \times10^{22}$ \columndensity\ (Fig. \ref{fig:xray}).
The Milky Way line of sight reddening is $E(B-V)=0.0285$ and that through the host galaxy  is
$E(B-V)=0.016$ \citep{sa20}. These values correspond to their respective column densities of $N_{H,\rm MW}=
 1.64\times10^{20}$   \columndensity\ and $N_{H,\rm host}=0.92\times 10^{20} $  \columndensity, using \citet{sf11} recalibration of
 the \citet{sfd98} extinction map. To derive $N_H$ we use $N_H=E(B-V) \times 5.8 \times 10^{21}$ \columndensity. However, this relation is an empirical relation and has uncertainties \citep[e.g.][]{gbp12,liszt19}.
 Combining the column density due to the Milky Way and the host galaxy gives $2.56\times10^{20}$   \columndensity, which has negligible effect on the 
 best fit column density which is two orders of magnitude higher.
The  excess  column density comes from the CSM, i.e.  
$N_{H,\rm CSM} = \sim 1.10 \times 10^{22}$ \columndensity.

\begin{figure}
 \centering
\includegraphics*[width=0.5\textwidth]{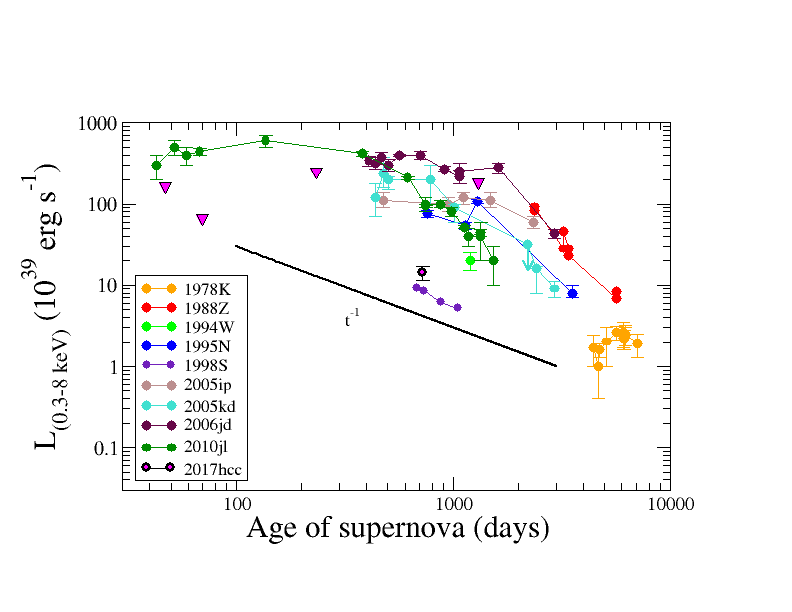}
\caption{\small The 0.3--8\,keV luminosities of X-ray detected  SNe IIn.  The {\it Chandra} \snhcc\ detection  is marked with a big magenta circle and
 the {\it Swift} upper limits with inverted triangles.  Reproduced from \citet{chandra18}. 
\label{fig:comparison}}
\end{figure}

The $0.3 -10.0$\,keV unabsorbed flux is $f_{0.3-10\,\rm keV}=(2.87\pm0.53) \times 10^{-14}$\,\flux, corresponding to  a luminosity of  $L_{0.3-10\,\rm keV}=(1.94\pm0.43)\times 10^{40}$\,\luminosity. { This value does not change significantly even if we use an X-ray emitting plasma 
temperature of 20\,keV  based on the temperature observed in SN 2010jl \citep{ofek+14, chandra+15}. } In Fig. \ref{fig:comparison}, we plot the \snhcc\ X-ray luminosity along
with other well-observed SNe IIn. Other than SN 1978K and SN 1998S, detected SNe IIn are brighter than \snhcc.
The \snhcc\ luminosity is  comparable to that of SN 1998S at the same age. We discuss the possible reasons for low X-ray luminosity in \S \ref{sec:discussion}.

\subsection{IR analysis}

\begin{figure}
\begin{center}
\includegraphics[width=0.49\textwidth,angle=0]{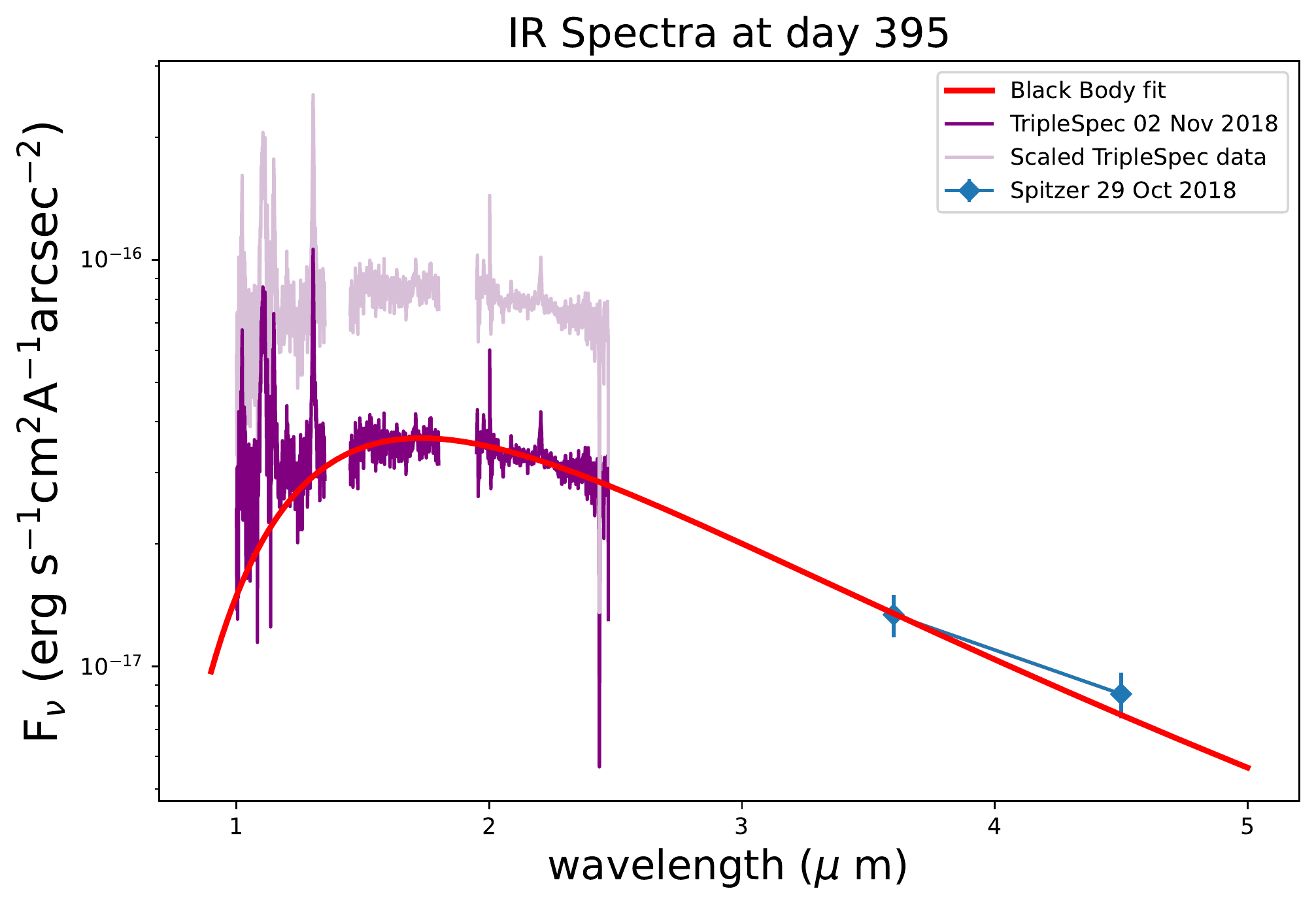}
\caption{The spectra on day $\sim 395$ combined with TripleSpec and Spitzer data. 
The light purple color shows original spectrum, whereas the purple spectrum has been shifted upwards to match the blackbody curve passing through the 3.6 $\mu$m \spitzer\  data point.
\label{fig:tedi}}
\end{center}
\end{figure}

\snhcc\ is extremely bright at \spitzer\ IRAC wavelengths. 
The  4.6 $\mu$m absolute magnitude reached $-21.7$ mag after a year,
 which corresponds to a peak IR luminosity  $\sim 10^{42}$ \luminosity.
 The late time emission
 is dominated by the IR emission \citep{sa20}, as is frequently the case for SNe IIn (see \S \ref{sec:intro}).  We estimate the temporal evolution $\beta $ (where $\beta$ is defined as $F_\nu (t)  \propto t^\beta$)  and between \spitzer\ flux densities ($F_\nu(t)$)  in $\mu$Jy (Fig. \ref{fig:spitzer}).
 The temporal evolution between epochs 1 and 2, at 3.6 and 4.5 $\mu$m, are  $\beta=-0.46\pm0.16$ and $\beta= -0.17\pm0.15$, respectively.
 The temporal indicies between the epochs 2 and  3 at the two frequencies are  
$\beta=-1.10\pm0.21$ and $\beta=-0.76\pm0.20$.    
The spectral indices $\alpha$  (defined as $F_\nu \propto \nu^\alpha$) between the two wavelengths at each epoch are 
$\alpha= -0.01\pm 0.25$,  $\alpha=0.47\pm 0.26$ and $\alpha=0.94\pm0.31$, respectively. The slow decline of the IR emission, combined with the fast optical decline \citep{kumar+19,sa20},
has important implications for the origin of the dust (\S \ref{sec:dust}).

The   TripleSpec data at five  epochs: 2017 Dec 5 (d 65), 2018 Jan 6 (d 97), Sep 18 (d 352), Nov 02 (d 397) and 2019 Jan 17 (d 473) are plotted in Fig. \ref{fig:triplespec}.
While the early data are dominated by shorter wavelength emission, the later data reveal higher flux densities at longer wavelengths.  A growing IR excess
over the photospheric emission  develops over time. 
 We combine the  measurements of \spitzer\ and TripleSpec on   2018 Oct 29 and 2018 Nov 2,
respectively, and fit a simple blackbody model to the combined spectrum.  The IR spectra using these data are plotted in 
Fig. \ref{fig:tedi}. 
However, the calibration of the TripleSpec data has large uncertainties and hence the  flux should not be considered absolute.  Ideally it should be scaled to near-IR photometry.
  We have  varied the scaling of the TripleSpec spectrum manually, and tried to shift to the blackbody curve passing through the Spitzer 3.6 $\mu$m data point. 
While the blackbody curve with $T\approx 1600$\,K fits the data reasonably well, there seems to be a slight excess of the 4.5 $\mu$m flux. However, the  significance of the fit is low due to the calibration uncertainties and  the fact that the scaled TripleSpec spectrum does not seem to go through both the Spitzer data points and thats causes the uncertainty in the scaling factor. Hence we do not use the parameters derived from these fits.

\begin{figure*}
\begin{center}
\includegraphics[width=0.86\textwidth,angle=0]{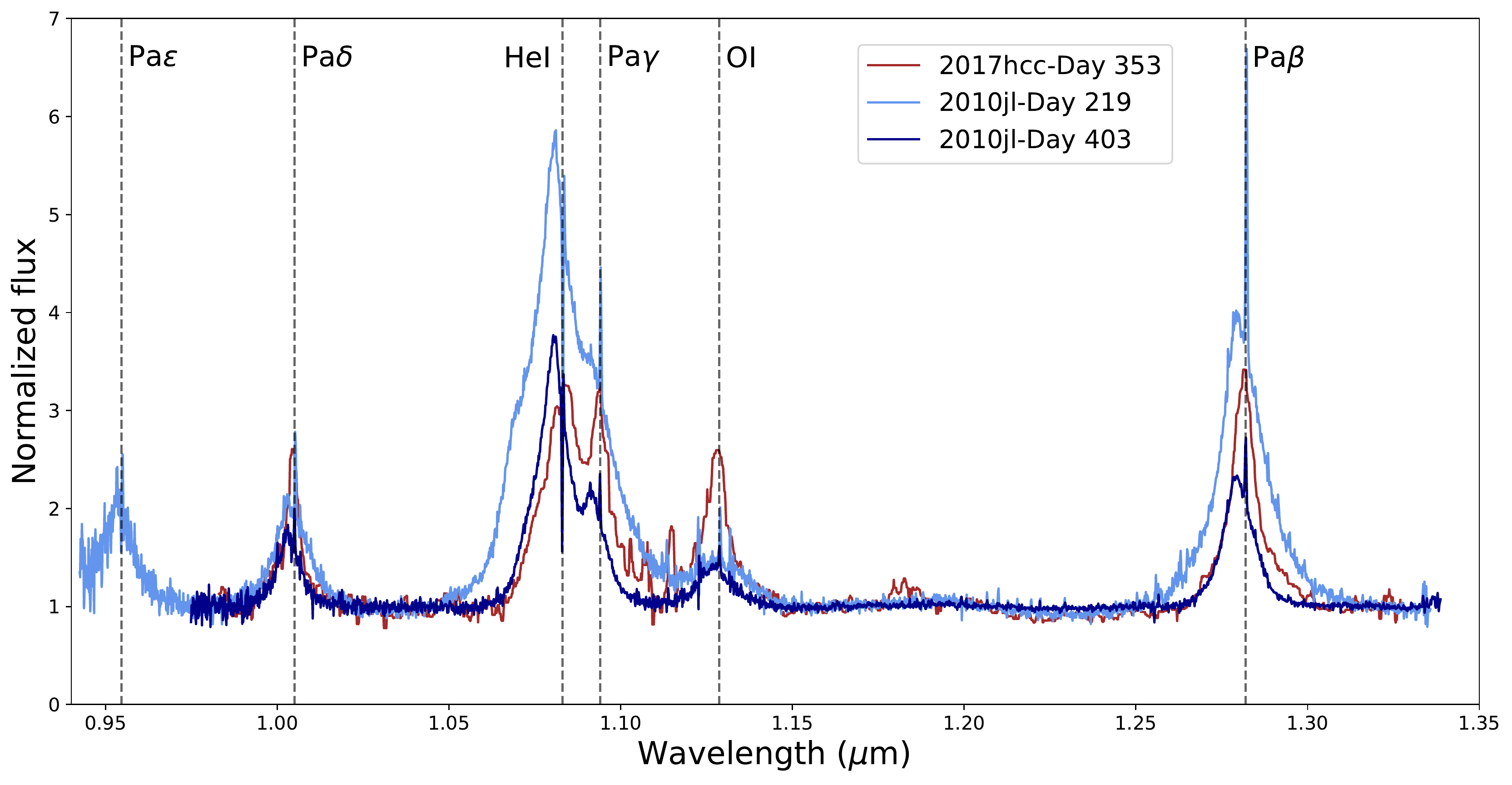}
\includegraphics[width=0.405\textwidth,angle=0]{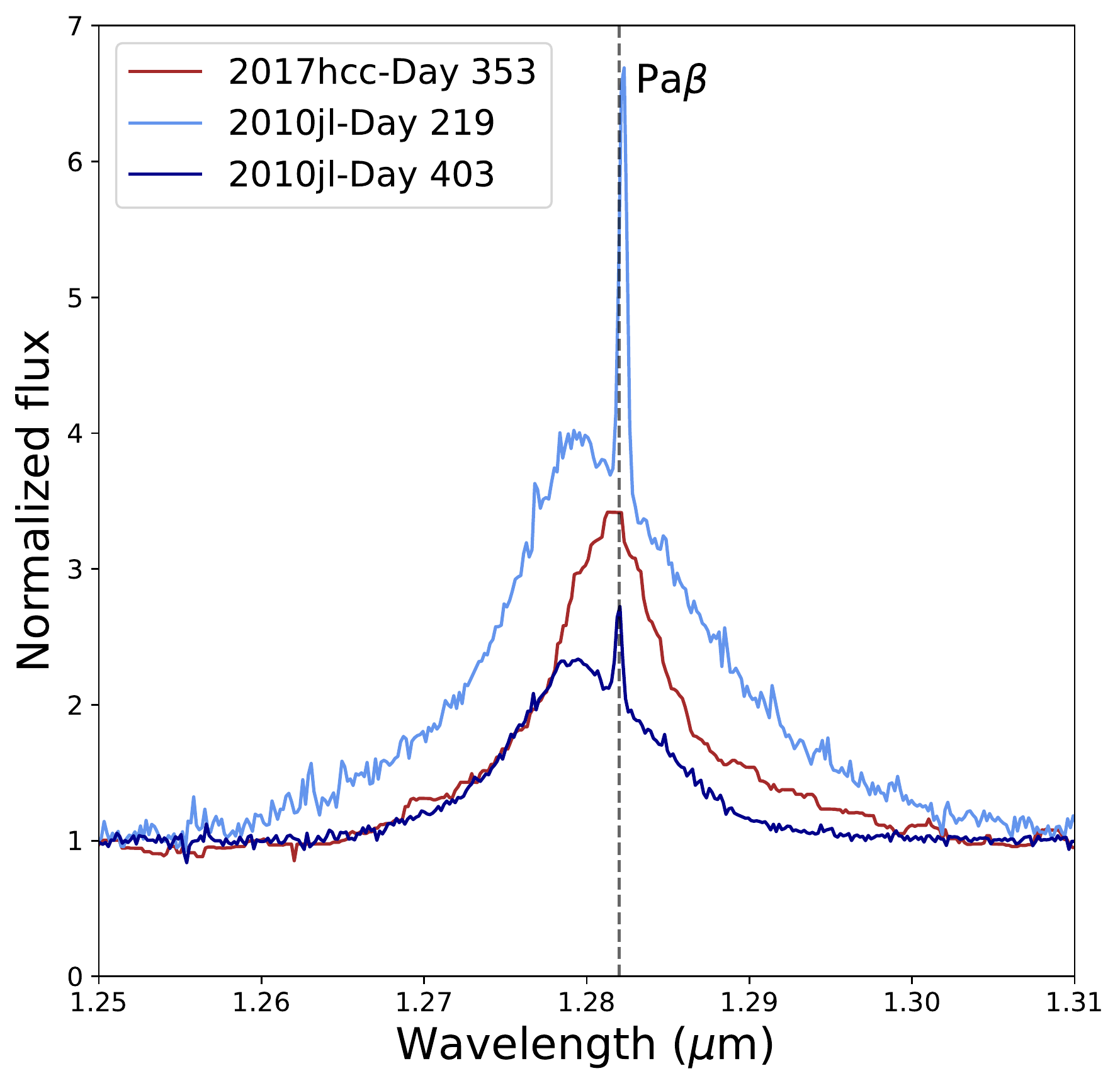}
\includegraphics[width=0.565\textwidth,angle=0]{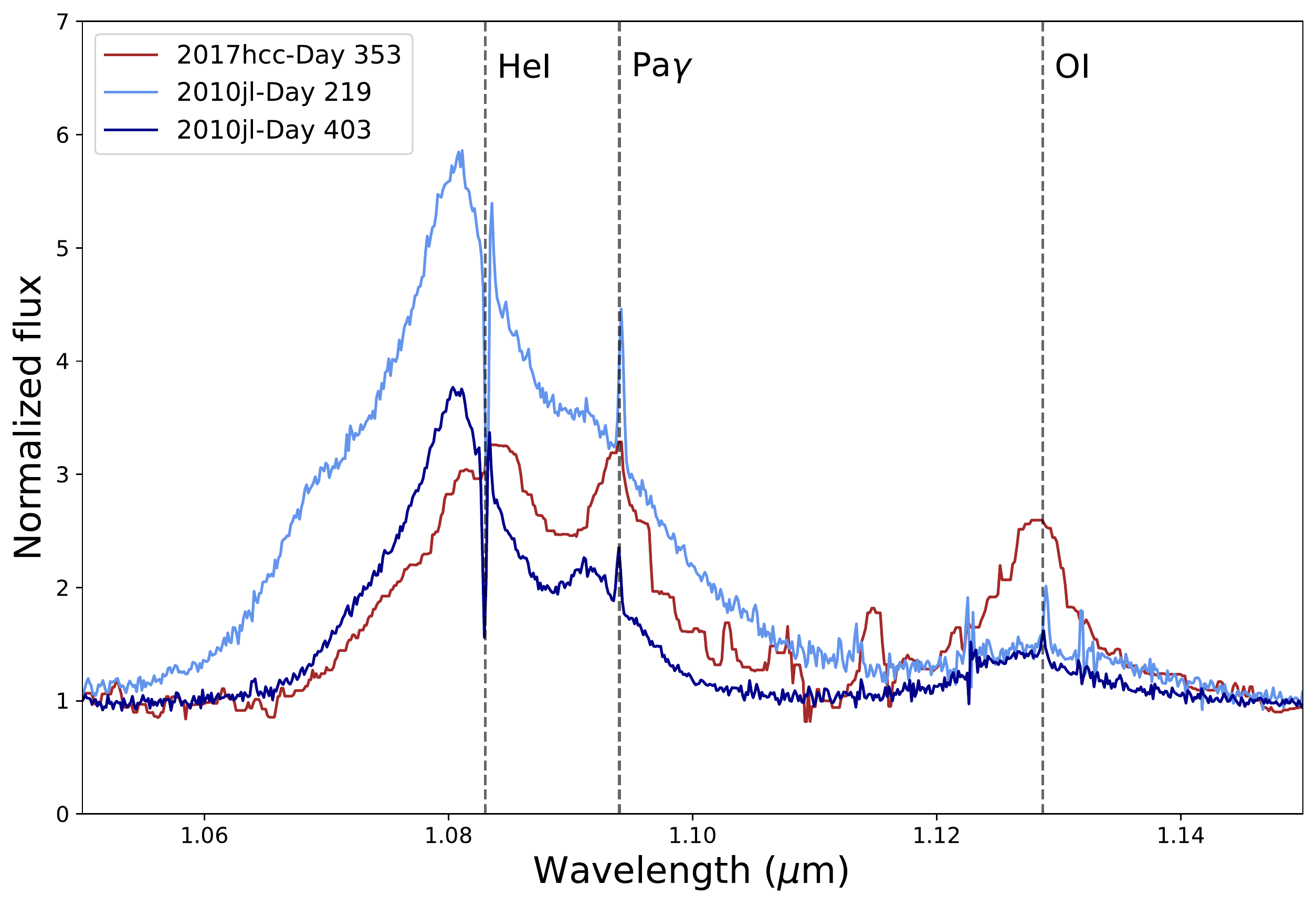}
\caption{Comparison of near IR spectra of \snhcc\ \citep{borish+15} with those of SN 2010jl. Spectra are normalized to the continuum level.   The \snhcc\ spectrum is on day 353 (brown lines, corresponding to Triplesec data on 2018 Sep 18 (MJD 58379)).
SN 2010jn spectra are on day 219 (light blue lines) and day 403 (dark blue lines) Top panel: The full spectra.  Bottom two panels  show zoomed plots for O I,   
Pa $\beta$, Paschen $\gamma$ and He I lines.}
\label{fig:irspectra}
\end{center}
\end{figure*}

We  compare  the \snhcc\ near IR TripleSpec spectra with those of SN 2010jl in  Figure \ref{fig:irspectra}.
The two  SNe have qualitatively similar characteristics but there 
are  significant differences. In \snhcc, the O I (1.129 $\mu$m)   line is much stronger relative to Pa$\beta$   (1.282 $\mu$m) than  in SN 20201jl, while the He I (1.083 $\mu$m)  line is weaker relative to Pa$\beta$.
 The H lines show some asymmetry towards the blue side  in \snhcc, This is  seen visually as a greater area under the lines to the blue side of the central wavelength vs the red side.
But the asymmetry is less than that in SN 2010jl. 
This could be due to  dust formation in the post-shock shell and
in the SN ejecta, which progressively obscures the redshifted part. Any associated CS dust was presumably evaporated by the SN.  
 Narrow lines like those observed in SN 2010jl are present in \snhcc\  \citep{sa20}, but are not resolved here.
The width of the expected narrow lines is $\sim 50$ \kms, whereas the TripleSpec resolution is $\sim 100$ \kms.
The estimated wind velocity is 100 \kms  for SN 2010jl  \citep{fransson+14},  but is $40 - 50$ \kms  for SN 2017hcc  \citep{sa20}

\subsection{Radio Analysis}

\begin{figure}
\begin{center}
\includegraphics[width=0.49\textwidth,angle=0]{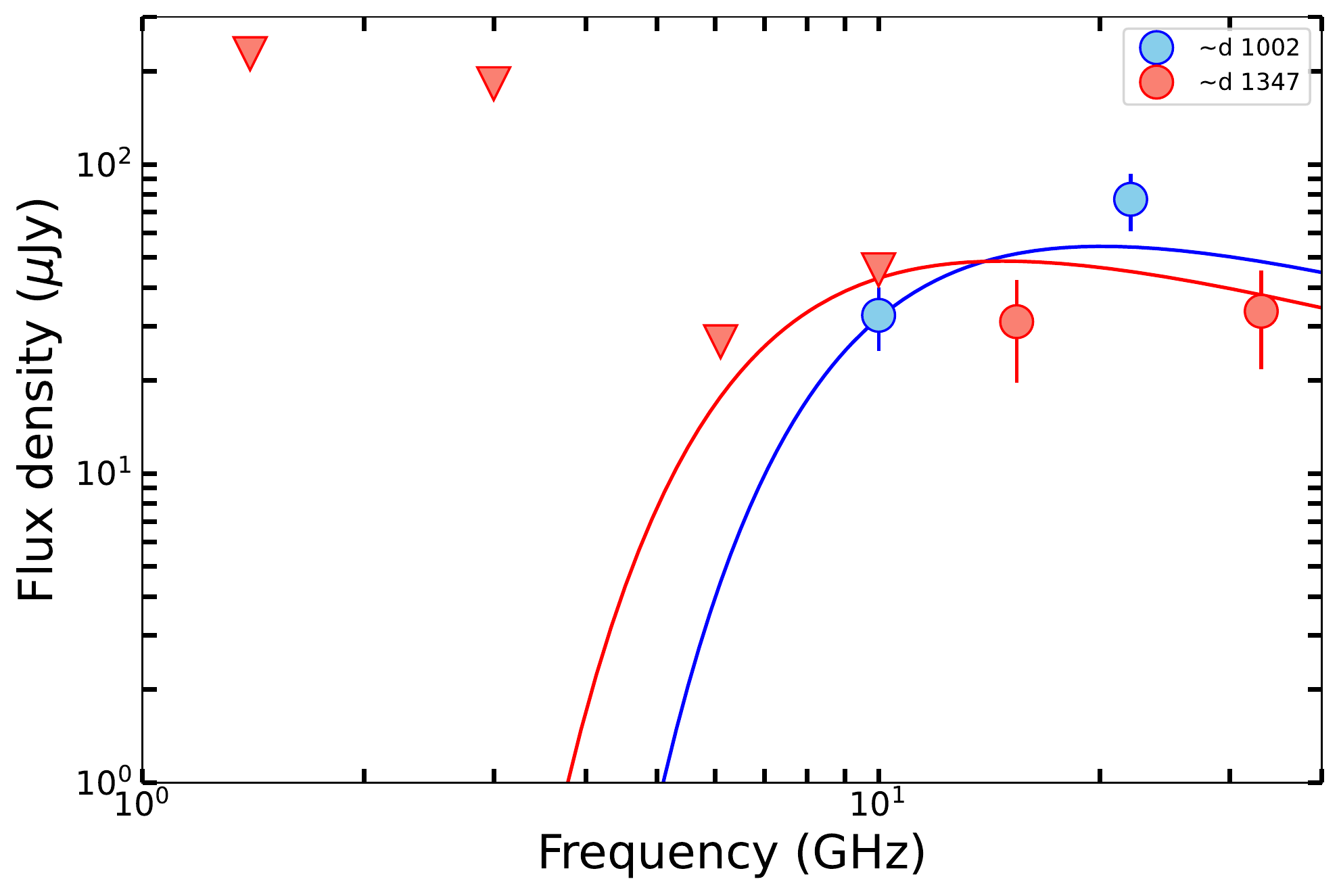}
\caption{Radio spectra of \snhcc\ with the VLA at two epochs.  The spectra on day 1000 is in blue color and the one on day 1347 is in red color. The triangles represent upper limits. 
A simple FFA model is fit to the data.
 \label{fig:radio}}
\end{center}
\end{figure}

The VLA measurements show  faint detections  ($\le5\,\sigma$, as mentioned in Table \ref{tab:radio} and Fig. \ref{fig:radio}).   The peak
spectral  luminosity is $5.2\times10^{26}$ \spectralluminosity, which  is towards the lower end of the SNe IIn luminosities \citep{chandra18}. 
The day 1000   spectral index between these frequencies is $+1.1$. This indicates that  the supernova is
still moderately  optically thick.  
Such behavior has been observed in the early evolution of other SNe IIn such as SN 1986J which showed
$\alpha  \sim +1.5$ in its rise to maximum \citep{weiler+90}
and has been attributed to clumpiness in the emitting region \citep{weiler+90}.
 Our radio data taken a year later   show a higher radio luminosity. The spectral indices between the detected points 
are flat, indicating that the radio emission  is near its  transition from optically thick to thin regime.
Due to the small number of data points, we fit the simple free-free absorption (FFA) model used in \citet{chandra+12b}  assuming an optically thin 
spectral index of $-0.6$. The best fits at the two epochs are shown in Fig. \ref{fig:radio}. 
We also attempt to  fit the synchrotron self-absorption model (SSA) and from the peak we estimate the SSA velocity using the
formulation in \citet{chevalier98}. This velocity is around $\sim 150$ \kms, which  is an order of magnitude smaller than the ejecta velocity. This rules out
SSA  as a significant absorption mechanism.

\section{Discussion}
\label{sec:discussion}

\subsection{Shock breakout in presupernova mass-loss} \label{sec:discussion1}


With a peak bolometric luminosity of $1.3 \times 10^{44}$ erg s$^{-1}$ reached in $\sim 30$ days and an absolute magnitude of $-20.7$ mag \citep{prieto+17}, \snhcc\ is at the boundary of 
superluminous  supernovae with a  two orders of magnitude larger luminosity than normal core-collapse supernovae.
In addition, strong narrow lines are present \citep{sa20}, indicating that dense outflowing circumstellar gas surrounds the supernova.
If this gas is sufficiently  optically thick, the shock breakout radiation from the supernova will occur after the interaction shell has moved to the mass-loss region. Quantitatively,   if the wind optical depth is $\tau_{\rm w}$ and $v_{\rm sh}$ is the shock speed, then for cases where $\tau_{\rm w} > c/v_{\rm sh}$, the SN shock  breaks out in the wind.
Such a  breakout through the
thick wind may extend long enough, lasting days to months,  and may account for
the peak SN luminosity of  \snhcc\  \citep{ci11}.

We use the approximate formulation of \citet{ci11} for shock breakout in a wind.
  An initial assumption for this is that the mass-loss density profile is that for a steady wind, $\rho = D r^{-2}$, where $D=\dot M/(4\pi R^2 v_{\rm w})$.
 As per their formulation, there are two length scales that are important. One, $R_{\rm d}$, is the radius at which the diffusion time equals the expansion time, so photons can move out through the wind.
  We have  $R_{\rm d}= \kappa D v_{\rm sh}/c$, where $\kappa$ is the opacity. 
  The other important length scale is $R_{\rm w}$ where the wind density cuts off. 
The character of the shock breakout is expected to depend on whether $R_{\rm w}>R_{\rm d}$, when the diffusion is important,  or vice versa for which the diffusion time scales are large \citep{ci11}.
  The indications are that  \snhcc\ is in the case $R_{\rm w}>R_{\rm d}$. This is indicated by
  the presence of bright infrared emission to late times (2 years) implying that the mass loss region extends to large radius.
  Also, the narrow line emission in optical bands \citep{sa20}  shows a dense mass loss region.
  We scale the parameter $D$  to $D_*$ such that 
$D_* = D/(5\times 10^{16}$ g cm$^{-1}$) corresponding to a mass loss rate of  $0.1 M_{\odot}$ yr$^{-1}$ and wind velocity of 100 km s$^{-1}$. Here $D$ is in g cm$^{-1}$ and   $D_*$  is a dimensionless quantity.
As mentioned earlier, in \snhcc\ the rise time can be approximated by the rise time of the bolometric light curve, 30 days \citep{prieto+17}, so $D_*\approx 4.5$.
  For the wind velocity of 45 km s$^{-1}$, we get $\dot M =0.1 M_{\odot}$ yr$^{-1}$ which should apply near the time of maximum light.
  
  At shock breakout, a radiation dominated shock can no longer be maintained and a cold dense shell (CDS) forms at the shock interface.
 A simple model for a supernova density profile, used by \citet{ci11}, is an inner region with a flat density surrounded by a steep $\rho \propto r^{-7}$ profile.
 For typical parameters, the interaction region between the ejecta and the CSM is in the steep power law region.
 For normal supernova parameters, Equation (6) in \citet{ci11} shows that the breakout luminosity can be close to that observed in \snhcc\ near bolometric peak.

  In the above we use 4000 km s$^{-1}$ for the shock velocity, as indicated in the X-ray observations of SN 2010jl \citep{chandra+15}
 and SN 2014C \citep{margutti+17}.
Here we discuss  the shock velocity.
\citet{sa20} determine $v_{\rm sh}$ from the emission line profiles of H$\alpha$ and H$\beta$.
They find that the line profiles can be fit by 2 Gaussian components, one with a FWHM of 4000 km s$^{-1} $
and one with  1100 km s$^{-1}$, which they identify with emission from the freely expanding ejecta and
the CDS, respectively.
After allowance for a geometrical factor, \citet{sa20} use  1600 km s$^{-1}$ for $v_{\rm CDS}$, where $v_{\rm CDS}=v_{\rm sh}$ for a thin shell.
However, 
there is no clear physical reason for the emission from the CDS or ejecta to have a Gaussian profile with a  width corresponding to a physical parameter.
\citet{dessart+15} modeled the emission from SN 2010jl and found reasonable estimates for the line profile
of H$\alpha$ considering emission from the CDS with electron scattering and CDS velocity $\sim 3000$ km s$^{-1}$.
Also, in a spherically symmetric interaction, a large, unexpected deceleration would be required between
the ejecta and the CDS for the values mentioned by \citet{sa20}.
In the formation of a thin shell between ejecta with power law index $n$ and CSM with power law $s$, we have $R\propto t^{(n-3)/(n-s)}$ \citep{Chevalier82}
so that $v_{\rm ej}/v_{\rm CDS} =(n-s)/(n-3)$.  
For typical values $n=8$ and $s=2$ \citep[e.g.,][for SN 2010jl]{fransson+14}, we have $v_{\rm ej}/v_{\rm CDS} =1.2$, as opposed to $v_{\rm ej}/v_{\rm CDS} =2.50$ from the line observations \citep{sa20}.
The low value of $v_{\rm CDS} $ is not consistent with  the type of evolution seen in SN 2010jl during the first year if there is spherically symmetric expansion.
Other estimates for $v_{\rm CDS}$ can  be considered.
In the case of SN 2010jl, an X-ray observation at 2 yr with NuSTAR showed a temperature of 20 keV
\citep{ofek+14, chandra+15}, which corresponds to a shock velocity of 4,000 km s$^{-1}$.
A higher velocity would be expected at earlier times and there is possible evidence for a higher velocity
 from a feature that appears in the He I line in SN 2010jl \citep{borish+15,chandra+15}.
 We consider both small and large values for $v_{\rm CDS}$.

For the shock front to break out in the wind region, $\tau_{\rm w}>c/v_{\rm sh}$ is needed.
 Taking $R_{\rm b} $ to be the radius of the base of the wind, we have $\tau_w\approx 5\times 10^{16} k R_{\rm b}^{-1} D_*$, where $k$ is the opacity $\kappa$ in units of 0.34 cm$^2$g$^{-1}$ \citep{ci11}.
As discussed above, for $D_{*}=4.5$, this gives $\tau_w \approx 76 (R_{\rm b}/{10^{15}{\rm cm}})^{-1}$. 
Thus for reasonable parameters, the shock breakout does occur in the wind.


 One aspect of shock breakout is that the temperature should increase with rising luminosity, since the initial rise to maximum
luminosity should be primarily due to heating of the photosphere. This has not been seen in \snhcc\ \citep{kumar+19}.
However, we note that the early measurements for temperature are not available before the peak in the light curve.  

\subsection{Post-Breakout Interaction}

Once the diffusion time is less than the supernova age, the luminosity corresponds to the power generated at  the shock.
We have a luminosity $L= 4\pi R^2 \epsilon \frac{1}{2} \rho v_{\rm sh}^3$, where $R$ is the radius of the CDS and $\epsilon$
is an efficiency factor for the conversion of shock power to radiation \citep{chugai90}.
At early times, the cooling time is less than the age so the shock power is radiated and $\epsilon \approx 1$.
 With sufficient good quality data, the winds of SNe IIn are generally found to be not steady \citep{dwarkadas07}, but the assumption should lead to a
rough  estimate for $\dot M$, which, for a steady wind is:
\begin{equation}
\dot M =  2Lv_{\rm w}/\epsilon  v_{\rm sh} ^3
\label{eq:mdot}
\end{equation} 

The value of $v_{\rm w}$ can be estimated from high spectral resolution observations.
\cite{sa20} find $v_{\rm w}\approx (40-50)$ km s$^{-1}$.
To estimate the value of $\dot M$, we use the bolometric luminosity of  $\sim  10^{42}$ erg s$^{-1}$     
which is the luminosity of infrared dust emission at $1-2$ yr  (Table \ref{tab:ir}).
We now find  from Equation \ref{eq:mdot} $\dot M = 2\times 10^{-3} \epsilon^{-1} L_{42}(v_{\rm sh}$/ 4000  km  s$^{-1})^{-3}$ $ M_{\odot}$ yr$^{-1}$, where  $v_{w}=45$ \kms is assumed, and $L_{42}$
is the bolometric luminosity in the units of $10^{42}$ erg s$^{-1}$.
So we now have $D_*\approx 0.024$.
We note that the value of $v_{\rm w}$ is determined over the first 100 days \citep{sa20}; there could be a change at later times.
In addition, the value of the mass loss is  sensitive to the shock velocity. However, 
our results indicate that the mass loss density drops more rapidly with radius than in the steady ($r^{-2}$) case.

The cooling of the postshock gas is important for the value of $\epsilon$ in the above formula.  We now discuss the validity of cooling. Our discussion is similar to that of \citet{cf17}.
For electron temperature $T_{\rm e}  \gtrsim 2.6 \times 10^7$  K, which is expected  for the post-forward shock wave, free-free cooling dominates with a cooling rate
$\Lambda  = 1.0 \times  10^{-23} (T_e/10^7 \, {\rm K})^{0.5}$ erg s$^{-1}$  cm$^3$. 
Assuming isobaric cooling behind the
shock, the cooling time is $ t_{\rm cool} =5kT_e / n_H   \Lambda(T_e)$.
Using \citet{cf17} eqn. (18), the transition time at which the age equals the cooling time, for ejecta velocity 4000 \kms (indicated with $v_{\rm ej,4}$ in the formula below) is (for $n=8$)
\begin{equation}
t_{\rm tr}=342 D_* v_{\rm ej,4} ^{-3}{\rm  days}.
\end{equation}
which is  75 days for our preferred parameters.
This estimate would imply that the supernova was past the cooling phase at the time of the X-ray observation.
However, the uncertainties are large so some effect of cooling cannot be ruled out.
The estimates should be taken with caution, since the properties of the absorbing gas are uncertain because of early Compton heating and radiative cooling as well as recombination \citep{lf88}.

\subsection{Variable mass-loss rate}

The above treatment shows that the mass-loss rate obtained at $\sim 30$ days from the shock breakout
was $\sim 0.1$ \massloss, whereas at few hundred days, obtained from the IR measurements,  was $\sim 2 \times 10^{-3}$ \massloss.
In addition, we can also use the  timing of the peak radio
 flux density  to estimate the mass-loss rate at $\sim 1000$ d. 
 The mass-loss rate derived from our radio measurement assuming a CSM temperature of $10^4$ K is  $\sim 6.5 \times10^{-4}$ \massloss.
These epochs correspond to roughly $\sim 10$, $\sim$100 and $\sim$300 days before explosion and the mass loss rate changes roughly 
two to three orders of magnitude.  This indicates a variable mass-loss rate, with enhanced mass-loss event occurring a decade before the supernova explosion.

There is growing number of evidence that many core-collapse supernovae undergo similar enhancements during the last years before the
explosion. \citet{maeda+21} reported evidence of enhanced mass-loss in  nearby type Ic supernova SN 2020oi in their
Atacama Large Millimeter  Array (ALMA)  observations, which they interpreted as
 the pre-SN activity  likely driven by   the nuclear burning activities in the star's final moments. Similar conclusions were drawn for the famous supernova SN 2014C undergoing metamorphosis from a stripped envelope supernova to a Type IIn via their radio and X-ray observations \citep{anderson+17, margutti+17,brethauer+22}.
The flash spectroscopy of various supernovae have also revealed the similar conclusions.
A remarkable example is SN 2013fs, discovered mere 3 hrs after the explosion by the automated real-time discovery and classification pipeline of the intermediate Palomar Transient Factory (iPTF) survey, which  enabled spectroscopy measurements within 6 hours of discovery \citep{yaron+17}.
The early measurements revealed a dense CSM to be confined within $10^{15}$ cm indicating enhanced mass-loss ($10^{-3}$ \massloss)
in the last one year before explosion in an otherwise normal core-collpase supernova. This led authors to suggest that such pre-SN
activities may be common in exploding stars.
\citet{shivvers+15} examined the Keck HIRES spectrum of SN 1998S taken within a few days after the SN and found convincing evidence for  enhanced mass-loss rate in the last 15 years of star's life. These measurements combined with other published measurements indicated much smaller mass-loss 
rate at earlier times, arriving at the conclusion of this section.
SN 2020tlf was a normal Type IIP/IIL supernova, where flash spectroscopy soon after the discovery revealed  
evidence of photoionization of CSM confined within $10^{15}$ cm created at a mass-loss rate of $10^{-2}$ \massloss\  in a  10--12 $M_\odot$ red supergiant 
progenitor, whereas 3 orders of magnitude smaller mass-loss rate at larger distances \citep{jg+22}.
Similarly \citet{terreran+22} also found the  presence of dense and confined CSM indicating  a phase of enhanced mass-loss  in its final moments of the progenitor of
SN 2020pni. \citet{tartaglia+21}  presented the very early phase to nebular phase observations of Type II supernova SN 2017ahn, discovered just a day after explosion.
Their modeling indicated a complex CSM surrounding the progenitor star with evidence of enhanced mass-loss in the last moments of
likely massive yellow supergiant progenitor.
While for SNe IIn, LBV scenario has been favored for enhanced mass-loss rates just before the explosion, 
above examples favor the idea that many core-collapse supernovae of different flavors may experience enhanced mass-loss in their final years, which could be governed by somewhat common physical mechanisms  \citep{terreran+22}.


\subsection{Low X-ray luminosity}

The lack of bright X-rays in \snhcc\ is surprising, as the detected SNe IIn are generally brighter than other core collapse
supernovae \citep{chandra18}.
 A dense interaction is indicated in \snhcc\ at early epochs by the optical luminosity and electron scattering line profiles.   
X-ray emission is detected at day 700 but the luminosity of \snhcc\ is lower than that typically found for strongly interacting supernovae of Type IIn and is closer to the luminosity observed from SN 1998S (Fig. \ref{fig:comparison}).  
In \snhcc\, during the time of the \chandra\ observation, most of the shock power goes into the IR, which is $\sim 10^3$ times the X-ray emission at the same epoch.
SN 1998S had late IR emission, but at a level $\sim 10^{40}$ erg s$^{-1}$  \citep{pozzo+05} at 2 years, which is nearly
two orders of magnitude smaller than that of \snhcc; the lower luminosity
was consistent with the lower mass-loss rate of  $2\times10^{-4} M_{\odot}$ yr$^{-1}$ deduced from the radio and X-ray emission \citep{pooley+02}.

When the interaction is strong,  the forward shock front is initially radiative and evolves to a non-radiative phase \citep{ci12}. 
 In the cooling phase, the CDS  created by the interaction is subject to the nonlinear thin shell instability (NTSI), giving rise to a corrugated structure  \citep{vishniac94}.  
 Numerical simulations of the instability show that the  X-ray emission is suppressed  \citep{sm18,kee+14}.
The reason for  X-ray suppression is 
the CDS that forms. The hot post-shock gas cools radiatively towards lower temperature and pressure in the presence of the CDS. 
Much of the cool emitting gas is underpressured relative to the hot surrounding medium. 
This pressure difference between the hot shock and the cool dense shell robs  the hot gas of its thermal energy,  which is now radiated with high efficiency at lower temperature. 
Another factor is that the corrugation of the forward shock gives oblique shocks that are weaker than head on shocks.  
However, as discussed above, estimates of the cooling at the time of the X-ray detection indicates that the shockwave is
probably  not cooling. In this case, the NTSI is not a factor,
although the uncertainties and assumptions may allow the cooling scenario. 

\subsection{Late time excess of IR emission and origin of the dust}
\label{sec:dust}

  The  \spitzer\ IR photometric data  corresponds to peak IR luminosities reaching $\sim 10^{42}$  \luminosity. 
   The origin of the bright IR emission at late times is not clear.  
   The IR 
   evolution is rather flat, as opposed to a faster decline seen in the optical band \citep{kumar+19}, though at earlier times. 
   Usually  the IR excess  
   imply a rising relative contribution of the dust component to the bolometric luminosity as the optical light-curve fades.
 The late-time  high IR luminosity 
 and the systematic blue-shift in the line profiles \citep{sa20} suggest that \snhcc\ likely has contributions from two components: 1)  dust formation in the 
ejecta and the CDS, and 2) pre-existing dust beyond the evaporation radius, giving rise to IR emission.


   \citet{sa20} find   that \snhcc\ line profiles show a progressively increasing blueshift.  They reject the possibility of this feature being due to
occultation by the SN photosphere,
pre-shock acceleration of the CSM, or asymmetric CSM, and 
explain it to be arising due to absorption by  newly formed dust in the post-shock shell and then SN ejecta. 

 Based on our IR fits, we  can calculate the blackbody radius, $r_{\rm bb}$, which is the  minimum size of the dust emitting component, 
 using the  modified blackbody  expression given in \citet{fox+10}. Using the best
 fit values in Table \ref{tab:ir}, the bolometric luminosities assuming blackbody emission are $4.79\times10^{41}$ \luminosity,
   $3.73\times10^{41}$ \luminosity\ and  $2.70\times10^{41}$ \luminosity\ at epochs 393,
 568 and 770 days, respectively.  
  These luminosities imply that the values of $r_{\rm bb}$ are  $1.6\times10^{16}$ cm, $2 \times10^{16}$ cm and  $2.2\times10^{16}$ cm at these
  epochs, respectively. 
   The shock radii at these epochs are  $(0.5-1.3) \times10^{16}$,  $(0.8-2.0) \times10^{15}$ and $(1.1-2.7) \times10^{16}$ cm, respectively for
shock velocity ranging between $1600-4000$ \kms.  We note that these values have significant uncertainties arising from assumption of single graphite grain of size 1 $\mu$m and a single temperature, as well as due to uncertainties in the opacity measurements.
These findings  give interesting clues to the dust origin.  The shock can 
 typically destroy any pre-existing dust within the shock radius. 
 The shock radii are smaller  (or equal for higher ejecta velocity) than the blackbody  radii at each epoch. This implies that there could still be pre-existing dust that is beyond the outer shock front that can give the IR emission \citep{fox+11}, in addition to newly formed dust and can cause IR excess.


The dust luminosity of SNe IIn in general is higher than in other SNe, suggesting that  the dense CSM is important for dust formation in strongly interacting  SNe IIn \citep{tinyanont+16}. 
 However, most of the SNe IIn with a high IR flux were also associated with a high X-ray luminosity.  This is not the case with \snhcc.  Fig. \ref{fig:comparison}
shows the \snhcc\ X-ray luminosity 
with other well observed SNe IIn. Other than SN 1978K and SN 1998S, the detected SNe IIn are brighter than \snhcc.
 The luminosity of  \snhcc\ is  comparable to that of SN 1998S at the same age, which is known to be due to lower CS interaction. This is unlikely to be the case for \snhcc.

\subsection{Asymmetry in the CSM}

Asymmetry has been seen in many SNe IIn.  For example, the large X-ray luminosity of SN 2006jd implied a high density, but a small amount of photoabsorption in the spectrum implied a low density
\citep{chandra+12b}. These conflicting measurements suggested an asymmetric CSM.  This seems to be a general feature of interacting CSM, and may 
indirectly  support the luminous blue variable (LBV) progenitor scenario \citep{smith14}.
Direct evidence for asymmetry in SN 2017hcc comes from measurements of  high polarization of the photospheric emission \citep{mauerhan+17,kumar+19}.
The fact that the polarized fraction declined with time indicates that the asymmetry was in the CSM, not the supernova ejecta.

 Also,  \citet{sa20}  found, based on blueshifted line profiles, that the intrinsic emission-line profile from the fast SN ejecta is symmetric, suggesting that the underlying SN explosion itself was not highly aspherical, and  the polarization is likely related to the CSM.
Another line of argument regarding the highly asymmetric CSM in \snhcc\ suggests that the polarization measurements are close
 to maximum. 
 The emission at  maximum light in SNe IIn is dominated by  interaction of the ejecta with the CSM,  implying
  that the major source of intrinsic polarization at maximum is due to this interaction, particularly with an asymmetric CSM. 
  The luminosity from the interaction 
  becomes progressively weaker as the CSM around the SN evolves into an optically thin state, resulting in a decrease in polarization.
  In addition, \citet{sa20} find that even though optical depths of the CSM are quite high, the SN ejecta emerge by day 75, which requires a non-spherical geometry of the CSM.

    \citet{sa20}  noted that the line profiles during the first  $\sim 100$ days are remarkably symmetric, indicating that the CSM speed remains similar over a large range of radii  regardless of the changing 
   supernova luminosity. The observed asymmetry at later times  is mild and limited to low radial velocities.  Based on these two arguments, they suggested that the asymmetry is in the form of  axisymmetry and our line of sight is close to the polar region, which when viewed from the earth at an intermediate latitude will show asymmetry.

As the main power source for SNe IIn is the kinetic energy of the SN ejecta, a large fraction of the ejecta kinetic energy can be 
converted into radiation due to the high density CSM.
However, an axisymmetric CSM could lead to an  inefficient conversion of kinetic energy into radiation in some directions, causing a lower X-ray flux. 
A low column density can  be explained by the axisymmetry. Our radio measurements may also support this view. The 
radio flux is very faint which may indicate absorption of the flux.
Thus it is possible that the bright IR 
is arising from the same region, which may be more symmetric and is different from the region where low luminosity radio and X-rays arise \citep[see Fig. 14 of ][]{sa20}.
In this scenario,  the X-rays are completely absorbed in the region of high luminosity and the observed X-rays are from a region of weak interaction with 
relatively lower column density. 


\section{Summary and conclusions}
\label{sec:conclusions}

\snhcc\ had one of the highest bolometric luminosities for a SN IIn, but low X-ray and radio luminosities. 
 The peak bolometric luminosity
of $\sim 10^{44}$ ergs s$^{-1}$  \citep{prieto+17}  is likely to be due to CS interaction.   
During the first 100 days, there are only upper limits on the X-ray luminosity,
$<3\times 10^{40}$ ergs s$^{-1}$, so the ratio of X-ray to total luminosity is $<0.0003$.
These results are comparable to those observed in SN 2006gy \citep{ofek+07,smith+07}. 
\snhcc\ has properties that are consistent with shock breakout in a dense wind \citep{ci11}.

The optical/IR lines during the first 100 days show electron scattering profiles that require an electron scattering optical
depth of at least a few, which corresponds to a column density of at least $10^{24}$ cm$^{-2}$.
Our  early time \swift\ measurements result in a non-detection; hence, we are unable to constrain the column depth. It is possible that the early  X-rays were  absorbed due to the high column density. Shock breakout in a dense medium also results in suppression of X-ray emission. In addition, the non-linear thin-shell instability
due to a radiative forward shock   results in further suppression of the X-rays.

There is  a late time enhancement in the IR emission with
peak IR luminosities reaching $\sim 10^{42}$ erg\,s$^{-1}$. At the same epoch, the X-ray luminosity is two orders of magnitude lower. We interpret  the high  IR luminosity as 
likely due to contributions from new dust as well as old dust. 

We also show that \snhcc\ likely underwent a phase of enhanced mass-loss years before explosion. This has been seen in many SNe IIn and may indirectly
support an LBV scenario, though the fine tuning between an LBV undergoing enhanced mass-loss and SN explosion remains an important issue. Thus  \snhcc\ adds an
important input towards our understanding of highly interacting SNe, whose progenitors remain a mystery.

\section*{Acknowledgements}
We thank the referee for useful comments, which helped improve the manuscript.
RAC acknowledges support from     
NASA through Chandra
award number GO0-21054X issued by the Chandra X-ray Center and from NSF grant     AST-1814910.
PC acknowledges support of the Department of Atomic Energy, Government of India, under project no. 12-R\&D-TFR-5.02-0700.
The Chandra Center is operated by the Smithsonian Astrophysical Observatory
for and on behalf of NASA under contract NAS8-03060.
 The National Radio Astronomy Observatory is a facility of the National Science Foundation operated under cooperative agreement by Associated Universities, Inc.
This research has made use of data and/or software provided by the High Energy Astrophysics Science Archive Research Center (HEASARC), which is a service of the Astrophysics Science Division at NASA/GSFC.
This work is based [in part] on observations made with the Spitzer Space Telescope, which was operated by the Jet Propulsion Laboratory, California Institute of Technology under a contract with NASA. Support for this work was provided by NASA.
This work made use of data supplied by the UK Swift Science Data Centre at the University of Leicester.

\section*{Data Availability}
The data underlying this article will be shared upon a reasonable request to the corresponding author.




\bibliographystyle{mnras}
\bibliography{ms} 

\bsp	
\label{lastpage}
\end{document}